\documentclass[twocolumn]{autart}    

\usepackage{graphicx}  
\usepackage[cmex10]{amsmath}
\usepackage{amsfonts}
\usepackage{amssymb}
\usepackage{cases}
\usepackage{accents}
\usepackage{float}
\usepackage{mathrsfs}
\usepackage{xcolor}
\usepackage{epstopdf}
\usepackage{cite}
\newcommand{\rref}[1]{(\ref{#1})}

\newtheorem{theorem}{Theorem}
\newtheorem{remark}{Remark}
\newtheorem{definition}{Definition}
\newtheorem{proposition}{Proposition}

\begin{document}

\begin{frontmatter}

 \title{Stability Analysis and State-Feedback Stabilization of LPV Time-Delay Systems  with Piecewise Constant Parameters subject to Spontaneous Poissonian Jumps } 


\author{Muhammad Zakwan}\ead{muhammad.zakwan@epfl.ch}   

\address{Institute of Mechanical Engineering, Ecole Polytechnique F\'ed\'erale de Lausanne, 1015 Lausanne,  Switzerland}  

\begin{keyword}                           
LPV systems, output-feedback, stochastic hybrid systems, delay, Lyapunov-Krasovskii functional              
\end{keyword}                             

\begin{abstract}                           This paper discusses the stability analysis of linear parameter varying systems with a parameter-dependent delay where the parameters are assumed to be stochastic piecewise constants under spontaneous Poissonian jumps. Based on stochastic Lyapunov-Krasovskii functionals, we also provide sufficient synthesis conditions for the gain-scheduled state-feedback controller with memory in terms of parameter-dependent linear matrix inequalities (LMIs). Such synthesis conditions are computationally intractable due to
the presence of integral terms. However, we show that these LMIs can be equivalently represented by integral-free LMIs,
which are computationally tractable. Finally, we illustrate the applicability of the results through examples.
\end{abstract}

\end{frontmatter}

\section{Introduction}
Synthesizing gain-scheduled controllers has proved to be a plausible way to control real-world non-linear applications. Such plausibility is offered by the framework of LPV systems for the control of non-linear and time-varying systems \cite{BriatBook},\cite{mohammadpour2012control}.   
The myriad of applications pertaining to LPV systems includes: the automotive industry \cite{sename2013robust}, turbofan engines \cite{gilbert2010polynomial}, robotics \cite{kajiwara1999lpv}, and aerospace systems \cite{shin2000h}. Moreover, real-world applications are also often affected by time delays that can degrade the performance of the dynamical systems, or in the worst case; they can cause instability; see \cite{niculescu2001delay}. Time delays frequently appear in communication networks, mechanical systems, aircraft, robotized teleoperation, and many other domains; see \cite{chiasson2007applications}. Since time delays can also adversely affect the stability of LPV systems \cite{BriatBook}, \cite{zakwan2020distributedj}, it is natural to consider  LPV time-delay systems. For further details on  LPV time-delay systems, curious readers are referred to  \cite{wu2001lpv,zhang2002stability,zhang2005delay,briat2007lft,
briat2008parameter,briat2010memory}, and references therein.

We consider a class of  LPV time-delay systems where parameters and time delay are piecewise constants. Such a class can be considered a generalization of switched linear systems where the modes take values in an interval rather than a finite, countable set. The stability analysis of LPV systems with piecewise constant parameters has been presented in \cite{BRIAT201510}. Likewise, for LPV time-delay systems,  dwell-time based stability analysis, and clock-dependent gain-scheduled state-feedback controller synthesis by employing clock-dependent Lyapunov-Krasovskii functionals have been considered in \cite{zakwan2020Dwell}. However, in these works, the jumps in the parameter trajectories are assumed to be deterministic, and they follow several different notions of dwell-times, e.g., minimum dwell-time, range-dwell time, and average dwell-time. The paradigm has recently been shifted to consider LPV systems with stochastically evolving parameters, \cite{davis2018markov}, \cite{hosoe2018robust}. 
Such a framework corresponds to real scenarios where abrupt variations in the system structure can occur, for example, component failures, haphazard disturbances, varying interconnections between subsystems, and variations in the operating point of a non-linear plant. These systems are well-modeled by Markov jump linear systems (MJLS), which are a class of stochastic dynamical systems. Several papers have been devoted to the stability and control of MJLS, for instance, \cite{todorov2008output} and \cite{de2000output} and the references therein.  
LPV systems with piecewise constant parameters subject to spontaneous Poissonian jumps generalize the framework of MJLS with a finite or infinite countable set to the case where the mode takes values in an uncountable bounded set \cite{zakwan2019poisson}. The stability analysis of such systems is proposed in \cite{briat2018stability}, where a state feedback controller is designed by employing a bounded real lemma to ensure satisfactory $\mathcal{L}_2$ performance. On the same principles, a dynamic $\mathcal{L}_2$ output feedback controller for the same class of systems is reported in \cite{zakwan2019poisson}. 

In this paper, we consider LPV time-delay systems with piecewise constant parameters subject to spontaneous Poissonian jumps. This framework generalizes the notion of time-delay MJLS with finite/countable modes to an uncountable bounded set.  
The jumps in the parameter trajectories are assumed to be spontaneous and follow a Poissonian distribution, i.e., the time between two successive jumps is exponentially distributed. 
The resultant system is a piecewise deterministic Markov process, which is also known as a stochastic hybrid system, \cite{davis2018markov}, \cite{teel2014stability}. In our framework, the deterministic part consists of the state dynamics and the parameters of the
LPV system, whereas the stochastic part, considers a Markovian update rule for the parameters. 

The contributions of this paper are many folds. We provide two convex sufficient conditions for the mean-square stability with  $\mathcal{L}_2$ performance of LPV time-delay systems with piecewise constant parameters and stochastic delay subject to spontaneous Poissonian jumps. These conditions are based on a different parameterization of parameter-dependent stochastic Lyapunov-Krasovskii functionals that are reminiscences of that used in the context of deterministic
 LPV time-delay systems. Interestingly, these conditions combine the flow (deterministic) and the jump (stochastic) parts in a single one.
{The resulting conditions take the form of  parameter-dependent LMIs that have the peculiarity of involving integral terms; this renders the inequalities intractable for the synthesis of gain-scheduled state-feedback controllers.} Nevertheless, we borrowed a viable solution to this problem by employing an approach provided in \cite{briat2018stability}, \cite{zakwan2019poisson} that yields tractable semi-definite programs. Moreover,
we assume that parameter-dependent decision variables in LMIs are polynomial functions of parameters, which is quite a reasonable assumption because the parameters take values in a compact set. The polynomials can be used to approximate any continuous function in a compact set, \cite{briat2018stability}.

The stability conditions are then extended to  the gain-scheduled state-feedback controllers' synthesis problems that ensure $\mathcal{L}_2$ performance. We provide two synthesis conditions based on different Lyapunov-Krasovskii functionals. Our goal is to explore the consequences of different parametrizations of Lyapunov-Krasovskii functionals on the conservatism in the analysis and synthesis. According to the author's knowledge, the stability analysis and state-feedback controller design for such a class of systems have been discussed for the first time.

The general framework of stochastic hybrid LPV time-delay systems and related preliminaries are presented in Section 2. Stability analysis and state-feedback controller synthesis are provided in Section 3 and Section 4, respectively. The effectiveness of our approach is illustrated via examples in  Section 5. Finally, Section 6 provides concluding remarks and future directives.

 The notation will be simplified whenever no confusion can arise from the context.  Let $\mathbb R$ represent the set of real numbers, and let $\mathbb R^{n \times{m}}$ denote real  matrices of dimension $n \times{m}$. 
The cone of symmetric (positive definite) matrices
of dimension $n$ is denoted by $\mathbb{S}^n (\mathbb{S}^n_{>0})$. For $A$, $B \in \mathbb{S}^n$, the expression $A \prec (\preceq)B$ means that $A-B$ is negative
(semi)definite. For some square matrix $A$, we define $\textrm{Sym} [A] = A + A^T$. The asterisk symbol $(*)$ denotes the complex conjugate transpose of a matrix. The Lebesgue measure of a compact set $\mathscr{B}$ is denoted by $\mu(\mathscr{B})$.
$\mathbb{E}[\cdot]$ denotes the expectation and $\mathbb{P}[\cdot]$ is the probability measure. $||\cdot||$ defines the standard Euclidean norm in $\mathbb{R}^n$.

\section{System Description and Preliminaries}
We consider LPV time-delay systems with  piecewise constant parameters  subject to Poissonian jumps and parameter-dependent stochastic time delays. The stochastic hybrid dynamics is described by    
\begin{equation}
\begin{array}{rcl}
\label{equ:systemdef}
\dot{x}(t)&=&{A(\rho)}x(t)+ A_d(\rho)x(t-\tau(\rho)) \\
&&+ {B(\rho)}u(t)+E(\rho)w(t)\\ 
z(t)&=&{C(\rho)}x(t)+ C_d(\rho)x(t-\tau(\rho))
\\
&& +{D(\rho)}u(t)+F(\rho)w(t) \\
x(t) &=& \phi(t), \; \forall t \in [-h,0] \; ,
\end{array} 
\end{equation}
where $x\in \mathbb{R}^{n}$, $w\in \mathbb{R}^{n_w}$, $u\in\mathbb{R}^{n_u}$, and $z\in \mathbb{R}^{n_z}$ are the state, the exogenous input, the control input, and the regulated output, respectively. The parameter vector $\rho(t)$ is piecewise constant (i.e., $\dot{\rho} = 0$ between the jumps) and randomly changes its values with a finite jump intensity. We define $\rho(t)$ as 
\begin{equation}
\label{equ:2}
\mathbb{P} \left[ \rho(t + \Delta t) \in \mathcal{B}|\rho(t) = \rho \right] = \kappa(\rho, \mathcal{B})\Delta t + o(\Delta t) \;,
\end{equation}
where $\rho \in \mathscr{B}$, $\mathcal{B} \subset \mathscr{B}-{\rho}$ is measurable, and  $\kappa : \mathscr{B} \times \mathscr{B} \rightarrow \mathbb{R}_{\ge 0}$ is the \emph{instantaneous jump rate} such that $\rho \mapsto \kappa(\rho, A)$ is measurable and $A 
\mapsto \kappa(\rho, A)$ is a positive measure. Particularly, $\kappa(\rho, d\theta)$ are transition rates and $\bar{\lambda}(\rho) = \int_{\mathscr{B}} \kappa(\rho, d\theta)$ are intensities.
{Since 
the evolution of $x(t)$ depends on $x(s)$, for $t - h \leq s \leq t$,
$(x(t),\rho(t))_{t \geq 0}$ is not a Markov process. To cast the system described by \rref{equ:systemdef} and \rref{equ:2} into the
framework of Markov systems, we define a translation operator $x_t(s) = x(t + s)$, which takes values in some non-zero interval $s \in [-h, 0]$ such that $(x_t, \rho(t))_{t \geq 0}$ is a strong Markov process \cite{boukas2012deterministic}.} 
For the sake of simplicity, we assume that $\kappa(\rho, d\theta) = \lambda(\theta, \rho)d\theta$, where $\lambda$ is a polynomial function.
{Furthermore, $\tau(\rho) \in \mathscr{D}$ is the time-varying stochastic delay function, where $\mathscr{D} := \{\tau :\mathscr{B} \mapsto [0,h], h < \infty\}$ and the derivative of the delay is zero almost everywhere, except on a countable set of instants (at jumps) where it does not exist.} Finally, $\phi(t)$ is a vector-valued initial continuous function defined on the interval $[-h,0]$.
 
We introduce the following definitions and results  from \cite{briat2018stability}, which will be substantial in proving our main results.
%

\begin{definition}
The system \rref{equ:systemdef}-\rref{equ:2} is mean-square stable (MSS) if for an arbitrary initial condition $(x_{0},\rho_{0})$, we have $\mathbb{E}[||x(t)||^2_2] \rightarrow 0$ as $t \rightarrow \infty$.
\end{definition}

\begin{definition}
The $\mathcal{L}_2$-norm of a signal $w :[0, \ \infty) \mapsto \mathbb{R}^n$ is 
\begin{equation*}
||w||_{\mathcal{L}_2} =\left( \int_{0}^{\infty} \mathbb{E}[||w(s)||^2_2 ds] \right)^{1\over 2}.
\end{equation*}
If $||w||_{\mathcal{L}_2} < \infty $, then the signal is  of finite energy and $w \in \mathcal{L}_2$. 
\end{definition}
\begin{definition}[\cite{briat2018stability}]
The (stochastic) $\mathcal{L}_2$ gain of the affine map $\mathcal{L}_2 \ni w \mapsto z \in \mathcal{L}_2$ with $u  \equiv 0$ and $\ x(0) = 0$ induced by the system \rref{equ:systemdef}-\rref{equ:2} is 
\begin{equation*}
||w \mapsto z||_{\mathcal{L}_2 - \mathcal{L}_2} = \sup_{||w||_{\mathcal{L}_2} = 1} ||z||_{\mathcal{L}_2}\;. 
\end{equation*}
\end{definition}

\section{Stability and Performance Analysis}
In this section, we present two  convex stability conditions in terms of parameter-dependent LMIs for the stochastic hybrid system \rref{equ:systemdef}-\rref{equ:2} by employing two different parameterizations of stochastic Lyapunov-Krasovskii functional. 
\subsection{Stability Result Using Partially Parameter-dependent Lyapunov-Krasovskii Functional}
\begin{theorem}
\label{Thm:1}
For a given constant  $h \geq 0$, assume there exist matrix-valued functions $P:\mathscr{B} \mapsto \mathbb{S}^n_{>0}$, ${Z}:\mathscr{B} \times \mathscr{B} \mapsto \mathbb{S}^n$, constant matrices $Q \in \mathbb{S}^n_{>0}$, $R\in \mathbb{S}^n_{>0}$,  and a scalar $\gamma > 0$  such that the following equality
\begin{equation}
\label{equ:3a1}
\displaystyle{\int_{\mathscr{B}}} Z(\theta,\rho) d\theta = 0
\end{equation}
holds for all $\rho \in \mathscr{B}$, and the following LMI: 
\begin{equation}
\label{equ:middle1}
{\fontsize{9}{9} \selectfont
 \left[ 
\begin{matrix}
{\Lambda}_{11}(\theta,\rho) & {\Lambda}_{12}(\rho) & {P}(\rho) E(\rho) & C^T(\rho) &   h A^T(\rho) {R} \\ 
*  & {\Lambda}_{22} & 0 & C_d^T(\rho) &  h A_d^T(\rho){R} \\
* & * & -\gamma^2 I & F^T(\rho) &  h E^T(\rho) {R}  \\
* & * &  * & -I & 0  \\ 
* & * & * & * & - {R} 
\end{matrix}
\right] \prec 0 }
\end{equation}
with
\begin{equation*}
\begin{array}{lll}
\Lambda_{11}(\theta,\rho) &=& \emph{Sym}[P(\rho)A(\rho)] + \mu(\mathscr{B})\lambda(\theta,\rho)[P(\theta) - P(\rho)] \\
&&+ Z(\theta,\rho) + \delta(\rho)Q - R \\
\Lambda_{12}(\rho) &=& P(\rho)A_d(\rho) + R \\
\Lambda_{22} &=& -Q-R
\end{array}
\end{equation*}
holds for all $\theta,\rho \in \mathscr{B}$, where $\delta(\rho) = 1 + 2\bar{\lambda}(\rho)h$, $\bar{\lambda}(\rho) = {\int_{\mathscr{B}}}\lambda(\theta,\rho)d \theta$, and $\mu(\mathscr{B})$ is the Lebesgue measure of the set $\mathscr{B}$. Then, the system (\ref{equ:systemdef})-(\ref{equ:2}) is mean-square stable in the absence of disturbance $w$ and $u \equiv 0$. Moreover, the $\mathcal{L}_2$-gain of the map $w \mapsto z$ is at most $\gamma$.
\end{theorem}
{\it Proof:}
Consider the following parameter-dependent stochastic Lyapunov-Krasovskii functional: 
\begin{equation*}
\begin{array}{lll}
V(x_t,\rho) = V_1(x,\rho) + V_2(x_t,\rho) + V_{3}(x_t,\rho)\;,
\end{array}
\end{equation*}
where 
\begin{equation*}
\fontsize{9}{9} \selectfont
\begin{array}{lll}
V_1(x,\rho) &=& x^T(t)P(\rho)x(t) \\ 
V_2(x_t,\rho) &=&   \displaystyle{\int_{t-\tau(\rho)}^{t}} x^T(s) Q x(s) ds  \\ 
V_3(x_t,\rho) &=& h \displaystyle{\int_{-h}^0  \int_{t+s}^{t}} \dot{x}^T(\eta) {R} \dot{x}(\eta) d \eta ds \;.
\end{array}
\end{equation*} 
Employing the infinitesimal generator $\mathcal{A}(\cdot)$ from \cite{briat2018stability}, straightforward but tedious calculations yield 
\begin{equation}
\resizebox{0.5\textwidth}{!}{$
\label{equ:81}
\begin{array}{lll}
\mathcal{A}V_1 &=& x^T \left[ \textrm{Sym}[P(\rho)A(\rho)] + \displaystyle{\int_{\mathscr{B}} \lambda(\theta,\rho)[P(\theta) - P(\rho)]d\theta}  \right]x \\ 
&& + 2x^T P(\rho) A_d(\rho) x(t - \tau(\rho)) + x^T P(\rho) E(\rho) w \\ 
\mathcal{A}V_2 &\leq&  x^T Qx  - x^T(t - \tau(\rho)) Q x(t - \tau(\rho)) \\ 
&& + \displaystyle{\int_{\mathscr{B}}}\lambda(\theta,\rho) \left[ 
\int_{t - \tau(\theta)}^t x^T(s)Qx(s) ds \right.
\\ && \left. - \displaystyle{\int_{t - \tau(\rho)}^t} x^T(s)Qx(s) ds\right] d \theta   \\ 
\mathcal{A}V_3 &\leq& h^2 \dot{x}^T R \dot{x} - h \displaystyle{\int_{t-h}^t} \dot{x}^T(s) R \dot{x}(s) ds  \; . 
\end{array} $}
\end{equation}
For the stochastic formulation, we need to upper bound the integrals appearing in $\mathcal{A}V_2$. Consider the integral in $\mathcal{A}V_2$, i.e.
\begin{equation*}
\resizebox{0.5\textwidth}{!}{$
\begin{array}{rcl}
\displaystyle{\int_{\mathscr{B}}}\lambda(\theta,\rho) \left[ 
\int_{t - \tau(\theta)}^t x^T(s)Qx(s) ds 
 - \int_{t - \tau(\rho)}^t x^T(s)Qx(s)ds \right] d \theta  \;.
\end{array} $}
\end{equation*}
{We can rewrite the above term as  
\begin{equation*}
\begin{array}{rcl}
 \displaystyle{\int_{\mathscr{B}}\lambda(\theta,\rho) 
\int_{t - \tau(\theta)}^{t - \tau(\rho)}} x^T(s)Qx(s) ds  d \theta  \; .
\end{array}
\end{equation*}  
Following a similar methodology provided in \cite[Proposition 8.1.4]{BriatBook}, let us pose that $\tau(\theta) = \tau(\rho) + \epsilon_\tau(\rho,\theta)$, then we have  
\begin{equation}
\begin{array}{rcl}
&&\displaystyle{\int_{\mathscr{B}}\lambda(\theta,\rho) 
\int_{t - \tau(\rho) - \epsilon_\tau}^{t - \tau(\rho)}} x^T(s)Qx(s) ds  d \theta  \\ 
&\leq& \displaystyle{\int_{\mathscr{B}}\lambda(\theta,\rho) 
\int_{t - \tau(\rho) -\max\{0,\epsilon_\tau \}}^{t - \tau(\rho)-\min\{0,\epsilon_\tau \}}} x^T(s)Qx(s) ds  d \theta   \\
&\leq& \displaystyle{\int_{\mathscr{B}}\lambda(\theta,\rho) 
\int_{t - \tau(\rho) - h}^{t - \tau(\rho) + h}} x^T(s)Qx(s) ds  d \theta  \\ 
&\leq&  \displaystyle{ \int_{t - \tau(\rho) - h}^{t - \tau(\rho) + h} \left[ \int_{\mathscr{B}}\lambda(\theta,\rho)  d \theta \right]} x^T(s)Qx(s) ds  \\
&\leq& 2\bar{\lambda}(\rho)  h x^T(t) Q x(t) \; . 
\end{array}
\end{equation}
Therefore, the inequality $\mathcal{A}V_2$ can be written as
\begin{equation}
\label{equ:111}
\begin{array}{lll}
\mathcal{A}V_2 &\le& x^T \delta(\rho) Qx  - x^T(t - \tau(\rho))  Q x(t - \tau(\rho))  \;,
\end{array}
\end{equation} }
where $\delta(\rho) = 1 + 2\bar{\lambda}(\rho)h$ and $\bar{\lambda}(\rho) = \displaystyle{\int_{\mathscr{B}}}\lambda(\theta,\rho)d \theta$.

Now consider the inequality $\mathcal{A}V_3$, since $\tau(\rho) < h$, the inequality
\begin{equation*}
- h \displaystyle{\int_{t-h}^t} \dot{x}^T(s) R\dot{x}(s) ds  \leq - h \displaystyle{\int_{t-\tau(\rho)}^t} \dot{x}^T(s) R \dot{x}(s) ds
\end{equation*}
holds.

By employing Jensen's inequality \cite{gu2003stability},  we upper bound the integral as follows:
\begin{equation*}
\begin{array}{lll}
\mathcal{A}V_3 &\leq& \dot{x}^T h^2 R \dot{x} - h \displaystyle{\int_{t-\tau(\rho)}^t} \dot{x}^T(s) R \dot{x}(s) ds \\
\mathcal{A}V_3 &\le&  \dot{x}^T h^2 R \dot{x} \\
&& \hspace{-4mm}-\frac{h}{\tau(\rho)}  \big[ x(t) - x(t - \tau(\rho))\big]^T R \big[x(t) - x(t - \tau(\rho))\big].
\end{array}
\end{equation*}
Finally, bounding $-\frac{h}{\tau(\rho)}$ by $-1$, we get
\begin{equation}
\label{equ:151}
\begin{array}{lll}
\mathcal{A}V_3 &\le&  \dot{x}^T h^2 R \dot{x} \\
&& \hspace{-3mm}-\big[ x(t) - x(t - \tau(\rho))\big]^T R \big[x(t) - x(t - \tau(\rho))\big].
\end{array}
\end{equation}
Combining the first equality in \rref{equ:81}, the inequalities \rref{equ:111}, and \rref{equ:151} and letting $\mathcal{A}V \prec 0$, we reach at 
\begin{equation*}
\fontsize{8}{8} \selectfont
\begin{array}{lll}
 \xi^T \left[ 
\begin{matrix}
\Gamma_{11}(\theta,\rho) & P(\rho)A_d(\rho) + R & P(\rho)E(\rho) \\ 
* & -Q-R & 0 \\ 
* & * & 0 
\end{matrix}
\right]\xi 
+ h^2 \Psi^T R \Psi \prec 0  \;,
\end{array}
\end{equation*}
where 
\begin{equation*}
\begin{array}{lll}
\xi &=& col\left[ \begin{matrix}
x(t) & x(t - \tau(\rho)) & w(t)
\end{matrix}  \right] \\
\Gamma_{11}(\theta,\rho) &=& \textrm{Sym}[P(\rho)A(\rho)] + \displaystyle{\int_{\mathscr{B}} \lambda(\theta,\rho)[P(\theta) - P(\rho)]d\theta} \\
&& + \delta Q - R \\
\Psi &=& 
\left[ 
\begin{matrix}
A(\rho) & A_d(\rho) & E(\rho)
\end{matrix} \right] \;.
\end{array}
\end{equation*}
To prescribe the $\mathcal{L}_2$ performance, we employ \cite[Theorem 2]{briat2018stability}, and arrive at 
\begin{equation}
\label{equ:perf1}
\begin{array}{lll}
\mathbb{E}[ V(x_t,\rho) ] + \mathbb{E} \bigg[ \displaystyle{\int_{0}^t} (||z(s)||_2^2 - \gamma^2 ||w(s)||_2^2)ds\bigg] \le 0 \;.
\end{array}
\end{equation} 
Substituting $z(t)$ into the inequality \rref{equ:perf1} leads to 
$\xi^T(t) \Omega(\theta,\rho)  \xi(t) \prec 0$
with
\begin{equation}
\resizebox{0.42\textwidth}{!}{$
\label{equ:191}
\begin{array}{lll}
\Omega(\theta,\rho) = \left[ 
\begin{matrix}
\Omega_{11}(\theta,\rho) &  P(\rho)A_d(\rho) + R + C^T(\rho) C_d(\rho) \\ 
* & -Q-R+ C_d^T(\rho)C_d(\rho) \\ 
* & *  
\end{matrix}
\right.  \\ \hspace{20mm}
\left.
\begin{matrix}
 P(\rho)E(\rho) + C^T(\rho)F(\rho)  \\
  C_d^T(\rho)F(\rho)  \\
   -\gamma^2I + F^T(\rho)F(\rho)  
\end{matrix}
\right]
 + h^2 \Psi^T R \Psi \;,
\end{array} $}
\end{equation}
where 
\begin{equation*}
\resizebox{0.45\textwidth}{!}{$
\begin{array}{lll}
\Omega_{11}(\theta,\rho) &=& \textrm{Sym}[P(\rho)A(\rho)] + \displaystyle{\int_{\mathscr{B}} \lambda(\theta,\rho)[P(\theta) - P(\rho)]d\theta} 
\\ && + \delta(\rho) Q  
  - R + C^T(\rho)C(\rho) \;.
\end{array} $}
\end{equation*}
{The integral in \rref{equ:191} makes the derivation of convex design conditions a difficult task. To remedy this,  we employ a result from  \cite{peet2009positive}, which stipulates that the integral condition ${\int_{\mathscr{B}} \lambda(\theta,\rho)[P(\theta) - P(\rho)]d\theta}$ appearing in LMI \rref{equ:191} can be replaced by $\mu(\mathscr{B})\lambda(\rho,\theta)(P(\theta)-P(\rho)) + Z(\rho,\theta)$,  for all $\rho, \theta \in \mathscr{B}$, if and  only if there exists a matrix-valued function $Z:\mathscr{B} \times \mathscr{B} \mapsto \mathbb{S}^n$ such that  $\int_{\mathscr{B}}Z(\theta,\rho) d\theta = 0$ holds for $\rho \in \mathscr{B}$. Taking the standard Schur complement of \rref{equ:191} yields the LMI \rref{equ:middle1}. The proof is complete. \hfill $\square$ }

\subsection{Stability Result Using Fully Parameter-dependent Lyapunov-Krasovskii Functional}
We now introduce another stability result by employing a different parameterization of Lyapunov-Krasovskii functional than the one used in Theorem \ref{Thm:1}. In this variant, we consider parameter dependence in each decision variable appearing in Lyapunov-Krasovskii functional. This leads to the following result.
\begin{theorem}
\label{Thm:2}
For a given constant  $h \geq 0$, assume there exist matrix-valued functions $P:\mathscr{B} \mapsto \mathbb{S}^n_{>0}$, $Q:\mathscr{B} \mapsto \mathbb{S}^n_{>0}$, $R:\mathscr{B} \mapsto \mathbb{S}^n_{>0}$, $\mathcal{Q}:\mathscr{B} \mapsto \mathbb{S}^n_{>0}$, ${Z}:\mathscr{B} \times \mathscr{B} \mapsto \mathbb{S}^n$, and scalars $\hat{\lambda} > 0$, and $\gamma > 0$  such that following 
\begin{subequations}
\small
\begin{equation}
\label{equ:3a}
\displaystyle{\int_{\mathscr{B}}} Z(\theta,\rho) d\theta = 0
\end{equation}
\begin{equation}
\label{equ:3b}
\displaystyle{\int_{\mathscr{B}}} \lambda(\theta,\rho)Q(\theta) d \theta \leq \mathcal{Q}(\rho) 
\end{equation}
\begin{equation}
\label{equ:3c}
\displaystyle{\int_{\mathscr{B}}} \lambda(\theta,\rho) R(\theta) d \theta \leq \hat{\lambda} R(\rho)
\end{equation} 
\end{subequations}
hold for all $\rho \in \mathscr{B}$, and the following LMI: 
\begin{equation}
\label{equ:middle}
{\fontsize{9}{9} \selectfont
 \left[ 
\begin{matrix}
{\Lambda}_{11}(\theta,\rho) & {\Lambda}_{12}(\rho) & {P}(\rho) E(\rho) & C^T(\rho) &   \sqrt{\epsilon} A^T(\rho) {R}(\rho) \\ 
*  & {\Lambda}_{22}(\rho) & 0 & C_d^T(\rho) &  \sqrt{\epsilon} A_d^T(\rho){R}(\rho) \\
* & * & -\gamma^2 I & F^T(\rho) &  \sqrt{\epsilon} E^T(\rho) {R}(\rho)  \\
* & * &  * & -I & 0  \\ 
* & * & * & * & - {R}(\rho)  
\end{matrix}
\right] \prec 0 }
\end{equation}
with
\begin{equation*}
\begin{array}{lll}
\Lambda_{11}(\theta,\rho) &=& \emph{Sym}[P(\rho)A(\rho)] + \mu(\mathscr{B})\lambda(\theta,\rho)[P(\theta) - P(\rho)] \\
&&+ Z(\theta,\rho) + Q(\rho) + h \mathcal{Q} - R(\rho) \\
\Lambda_{12}(\rho) &=& P(\rho)A_d(\rho) + R(\rho) \\
\Lambda_{22}(\rho) &=& -Q(\rho)-R(\rho)
\end{array}
\end{equation*}
holds for all $\theta,\rho \in \mathscr{B}$, where $\epsilon = {h^2} + \hat{\lambda}\frac{h^3}{2}$, and $\mu(\mathscr{B})$ is the Lebesgue measure of the set $\mathscr{B}$. Then, the system (\ref{equ:systemdef})-(\ref{equ:2}) is mean-square stable in the absence of disturbance $w$ and $u \equiv 0$. Moreover, the $\mathcal{L}_2$-gain of the map $w \mapsto z$ is at most $\gamma$.
\end{theorem}
{\it Proof:}
We consider the following parameter-dependent stochastic Lyapunov-Krasovskii functional: 
\begin{equation*}
\begin{array}{lll}
V(x_t,\rho) = V_1(x,\rho) + V_2(x_t,\rho) + V_{3}(x_t,\rho)\;,
\end{array}
\end{equation*}
where 
\begin{equation*}
\fontsize{10}{10} \selectfont
\begin{array}{lll}
V_1(x,\rho) &=& x^T(t)P(\rho)x(t) \\ 
V_2(x_t,\rho) &=&   \displaystyle{\int_{t-\tau(\rho)}^{t}} x^T(s) Q(\rho) x(s) ds \\ &&+ \displaystyle{\int_{-h}^0  \int_{t+s}^{t}} x^T(\eta) \mathcal{Q}(\rho) x(\eta) d \eta ds \\ 
V_3(x_t,\rho) &=& h \displaystyle{\int_{-h}^0  \int_{t+s}^{t}} \dot{x}^T(\eta) {R}(\rho) \dot{x}(\eta) d \eta ds \\
&& + h \hat{\lambda} \displaystyle{\int_{-h}^0 \int_{s}^0  \int_{t+s}^{t}} \dot{x}^T(\nu) {R}(\rho) \dot{x}(\nu) d \nu d \eta ds\;.
\end{array}
\end{equation*} 
Employing the infinitesimal generator $\mathcal{A}(\cdot)$ from \cite{briat2018stability}, after  straightforward but involved calculations, we arrive at 
\begin{equation}
\label{equ:8}
\resizebox{0.48\textwidth}{!}{$
\begin{array}{lll}
\mathcal{A}V_1 &=& x^T \left[ \textrm{Sym}[P(\rho)A(\rho)] + \displaystyle{\int_{\mathscr{B}} \lambda(\theta,\rho)[P(\theta) - P(\rho)]d\theta}  \right]x \\ 
&& + 2x^T P(\rho) A_d(\rho) x(t - \tau(\rho)) + x^T P(\rho) E(\rho) w \\
\mathcal{A}V_2 &\leq&  x^T Q(\rho)x  - x^T(t - \tau(\rho)) Q(\rho) x(t - \tau(\rho)) \\ 
&& + \displaystyle{\int_{\mathscr{B}}}\lambda(\theta,\rho) \left[ 
\int_{t - \tau(\theta)}^t x^T(s)Q(\theta)x(s) ds \right.
\\ && \left. - \displaystyle{\int_{t - \tau(\rho)}^t} x^T(s)Q(\rho)x(s) ds\right] d \theta + h x^T \mathcal{Q}(\rho)x \\
&& - \displaystyle{\int_{t - h}^t} x^T(s) \mathcal{Q}(\rho)x(s) d s   \\
\mathcal{A}V_3 &\leq& h^2 \dot{x}^T R(\rho) \dot{x} - h \displaystyle{\int_{t-h}^t} \dot{x}^T(s) R(\rho) \dot{x}(s) ds \\ 
&&  + \displaystyle{\int_{\mathscr{B}}} \lambda(\theta,\rho) \left[ h \int_{-h}^0 \int_{t + s}^t \dot{x}^T(\eta) R(\theta) \dot{x}(\eta) d\eta ds \right. \\ 
&& -  \left.  h \displaystyle{\int_{-h}^0} \displaystyle{\int_{t + s}^t} \dot{x}^T(\eta) R(\rho) \dot{x}(\eta) d\eta ds \right] d \theta   \ \\ 
&& + \hat{\lambda}\frac{h^3}{2} \dot{x}^T R(\rho) \dot{x} - \hat{\lambda}h \displaystyle{\int_{-h}^0 \int_{t + s }^t } \dot{x}^T (\eta) R(\rho) \dot{x}(\eta) d \eta d s \; . 
\end{array} $}
\end{equation}
For the stochastic formulation, we need to upper bound the integrals appearing in $\mathcal{A}V_2$ and $\mathcal{A}V_3$. To this aim, let us consider the integral in $\mathcal{A}V_2$, i.e.
\begin{equation*}
\resizebox{0.48\textwidth}{!}{$
\begin{array}{lll}
 \displaystyle{\int_{\mathscr{B}}}\lambda(\theta,\rho) \left[ 
\int_{t - \tau(\theta)}^t x^T(s)Q(\theta)x(s) ds 
 - \int_{t - \tau(\rho)}^t x^T(s)Q(\rho)x(s)ds \right] d \theta  
\end{array} $}
\end{equation*}
{Since $Q(\rho) > 0 $, we drop the negative-definite term $- \int_{t - \tau(\rho)}^t x^T(s)Q(\rho)x(s)ds$ and get 
\begin{equation*}
\begin{array}{lll}
&&\displaystyle{\int_{\mathscr{B}}}\lambda(\theta,\rho) 
\int_{t - \tau(\theta)}^t x^T(s)Q(\theta)x(s) ds  d \theta  \\ 
&&\qquad \leq \displaystyle{\int_{\mathscr{B}}}
\int_{t - \tau(\theta)}^t \lambda(\theta,\rho)  x^T(s)Q(\theta)x(s) ds  d \theta  \\ 
&&\qquad = \displaystyle{\int_{\mathscr{B}}} \left[ 
\int_{t - h}^t \lambda(\theta,\rho)  x^T(s)Q(\theta)x(s) ds  \right. \\ 
&& \qquad \qquad \left. - \displaystyle{\int_{t - h}^{t - \tau(\theta)}} \lambda(\theta,\rho)  x^T(s)Q(\theta)x(s) ds \right] d\theta 
\end{array}
\end{equation*}  
Under the light of same argument, we drop the negative-definite term $-\int_{t - h}^{t - \tau(\theta)} \lambda(\theta,\rho)  x^T(s)Q(\theta)x(s) ds$ and employing the Fubini's Theorem yields
\begin{equation*}
\begin{array}{lll}
& \leq \displaystyle{\int_{t - h}^t}\int_{\mathscr{B}} x^T(s)\lambda(\theta,\rho)Q(\theta)x(s) d\theta ds \\
& \qquad  = \displaystyle{\int_{t-h}^t} x^T(s) \left[ \int_{\mathscr{B}} \lambda(\theta,\rho) Q(\theta) d\theta \right] x(s)  ds \; .
\end{array}
\end{equation*} 
Moreover, given that \rref{equ:3b} holds, we have  
\begin{equation*}
\resizebox{0.5\textwidth}{!}{$
\begin{array}{lll}
\displaystyle{\int_{t-h}^t} x^T(s) \left[ \int_{\mathscr{B}} \lambda(\theta,\rho) Q(\theta) d\theta \right] x(s)  ds \le  \displaystyle{\int_{t - h}^t} x^T(s) \mathcal{Q}(\rho)x(s) d s \; .
\end{array} $}
\end{equation*} 
Therefore, we can write $\mathcal{A}V_2$ as:
\begin{equation}
\label{equ:11}
\begin{array}{lll}
\mathcal{A}V_2 &\le& x^T Q(\rho)x  - x^T(t - \tau(\rho)) Q(\rho) x(t - \tau(\rho)) \\ 
&& + h x^T \mathcal{Q}(\rho)x \;.
\end{array}
\end{equation} }
Using the same arguments and employing \rref{equ:3c}, the following inequality holds for the integrals in  $\mathcal{A}{V_3}$
\begin{equation*}
\begin{array}{lll}
&& h \displaystyle{\int_{-h}^0 \int_{t + s}^t} \dot{x}^T(\eta) \left[ \int_{\mathscr{B}} \lambda(\theta,\rho)R(\theta) d\theta \right] \dot{x}(\eta) d\eta ds  \\ 
&&\qquad \qquad \le  \hat{\lambda}h \displaystyle{\int_{-h}^0 \int_{t + s }^t } \dot{x}^T (\eta) R(\rho) \dot{x}(\eta) d \eta d s.
 \end{array}
\end{equation*}
Hence, $\mathcal{A}V_3$ can be written as:
\begin{equation}
\label{equ:13}
\begin{array}{lll}
\mathcal{A}V_3 &\leq& \dot{x}^T \epsilon R(\rho) \dot{x} - h \displaystyle{\int_{t-h}^t} \dot{x}^T(s) R(\rho) \dot{x}(s) ds \;,
\end{array}
\end{equation}
where $\epsilon = {h^2} + \hat{\lambda}\frac{h^3}{2}$. Since $\tau(\rho) < h$, it holds that
\begin{equation*}
- h \displaystyle{\int_{t-h}^t} \dot{x}^T(s) R(\rho) \dot{x}(s) ds  \leq - h \displaystyle{\int_{t-\tau(\rho)}^t} \dot{x}^T(s) R(\rho) \dot{x}(s) ds \; .
\end{equation*}
By employing Jensen's inequality \cite{gu2003stability},  we upper bound the integral in \rref{equ:13} as:
\begin{equation*}
\begin{array}{lll}
\mathcal{A}V_3 &\leq& \dot{x}^T \epsilon R(\rho) \dot{x} - h \displaystyle{\int_{t-\tau(\rho)}^t} \dot{x}^T(s) R(\rho) \dot{x}(s) ds \\
\mathcal{A}V_3 &\le&  \dot{x}^T \epsilon R(\rho) \dot{x} \\
&& \hspace{-8mm}-\frac{h}{\tau(\rho)}  \big[ x(t) - x(t - \tau(\rho))\big]^T R(\rho) \big[x(t) - x(t - \tau(\rho))\big].
\end{array}
\end{equation*}
Finally, bounding $-\frac{h}{\tau(\rho)}$ by $-1$, we get
\begin{equation}
\label{equ:15}
\begin{array}{lll}
\mathcal{A}V_3 &\le&  \dot{x}^T \epsilon R(\rho) \dot{x} \\
&& \hspace{-3mm}-\big[ x(t) - x(t - \tau(\rho))\big]^T R(\rho) \big[x(t) - x(t - \tau(\rho))\big].
\end{array}
\end{equation}
Collecting the first equality in \rref{equ:8}, and inequalities \rref{equ:11}, \rref{equ:15} and letting $\mathcal{A}V \prec 0$, we have 
\begin{equation*}
\fontsize{8}{8} \selectfont
\begin{array}{lll}
 \xi^T \left[ 
\begin{matrix}
\Gamma_{11}(\theta,\rho) & P(\rho)A_d(\rho) + R(\rho) & P(\rho)E(\rho) \\ 
* & -Q(\rho)-R(\rho) & 0 \\ 
* & * & 0 
\end{matrix}
\right]\xi 
+ \epsilon \Psi^T R(\rho) \Psi \prec 0  \;,
\end{array}
\end{equation*}
where 
\begin{equation*}
\begin{array}{lll}
\xi &=& col\left[ \begin{matrix}
x(t) & x(t - \tau(\rho)) & w(t)
\end{matrix}  \right] \\
\Gamma_{11}(\theta,\rho) &=& \textrm{Sym}[P(\rho)A(\rho)] + \displaystyle{\int_{\mathscr{B}} \lambda(\theta,\rho)[P(\theta) - P(\rho)]d\theta} \\
&&+ Q(\rho) + h \mathcal{Q}(\rho) - R(\rho) \\
\Psi &=& 
\left[ 
\begin{matrix}
A(\rho) & A_d(\rho) & E(\rho)
\end{matrix} \right] \;.
\end{array}
\end{equation*}
We employ \cite[Theorem 2]{briat2018stability} to incorporate the $\mathcal{L}_2$ performance to reach the following inequality 
\begin{equation}
\label{equ:perf}
\begin{array}{lll}
\mathbb{E}[ V(x_t,\rho) ] + \mathbb{E} \bigg[ \displaystyle{\int_{0}^t} (||z(s)||_2^2 - \gamma^2 ||w(s)||_2^2)ds\bigg] \le 0 \;.
\end{array}
\end{equation} 
Substituting $z(t)$ into the inequality \rref{equ:perf} leads to the inequality
$\xi^T(t) \Omega(\theta,\rho)  \xi(t) \prec 0$
with
\begin{equation}
\resizebox{0.42\textwidth}{!}{$
\label{equ:19}
\begin{array}{lll}
\Omega(\theta,\rho) = \left[ 
\begin{matrix}
\Omega_{11}(\theta,\rho) &  P(\rho)A_d(\rho) + R(\rho) + C^T(\rho) C_d(\rho) \\ 
* & -Q(\rho)-R(\rho)+ C_d^T(\rho)C_d(\rho) \\ 
* & *  
\end{matrix}
\right.  \\ \hspace{20mm}
\left.
\begin{matrix}
 P(\rho)E(\rho) + C^T(\rho)F(\rho)  \\
  C_d^T(\rho)F(\rho)  \\
   -\gamma^2I + F^T(\rho)F(\rho)  
\end{matrix}
\right]
 + \epsilon \Psi^T R(\rho) \Psi \;,
\end{array} $}
\end{equation}
where 
\begin{equation*}
\resizebox{0.45\textwidth}{!}{$
\begin{array}{lll}
\Omega_{11}(\theta,\rho) &=& \textrm{Sym}[P(\rho)A(\rho)] + \displaystyle{\int_{\mathscr{B}} \lambda(\theta,\rho)[P(\theta) - P(\rho)]d\theta} 
\\ && + Q(\rho)  
 + h \mathcal{Q}(\rho) - R(\rho) + C^T(\rho)C(\rho) \;.
\end{array} $}
\end{equation*}
The rest of the proof follows from the same lines as in the proof of Theorem \ref{Thm:1}.

\subsection{Discussion}

{It is straightforward to see that under certain relaxations Theorem \ref{Thm:2} implies Theorem \ref{Thm:1}. For instance, consider the infinitesimal generator of Lyapunov-Krasovskii functional $\mathcal{A}V_2(x_t,\rho)$ in \rref{equ:13} without parameter dependence in the decision variables
\begin{equation*}
\begin{array}{lll}
\mathcal{A}V_2 &\leq&  x^T Qx  - x^T(t - \tau(\rho)) Q x(t - \tau(\rho)) \\ 
&& + \displaystyle{\int_{\mathscr{B}}}\lambda(\theta,\rho) \left[ 
\int_{t - \tau(\theta)}^t x^T(s)Qx(s) ds \right.
\\ && \left. - \displaystyle{\int_{t - \tau(\rho)}^t} x^T(s)Qx(s) ds\right] d \theta + h x^T \mathcal{Q}x \\
&& - \displaystyle{\int_{t - h}^t} x^T(s) \mathcal{Q}x(s) d s \; .
\end{array}
\end{equation*} 
Moreover, the above expression can be re-written as 
\begin{equation}
\label{equ:discus21}
\begin{array}{lll}
\mathcal{A}V_2 &\leq&  x^T Qx  - x^T(t - \tau(\rho)) Q x(t - \tau(\rho)) \\ 
&& + \displaystyle{\int_{\mathscr{B}}}\lambda(\theta,\rho) \left[ 
\int_{t - \tau(\theta)}^{t-\tau(\rho)} x^T(s)Qx(s) ds \right] d \theta  \\ 
&&+ h x^T \mathcal{Q}x - \displaystyle{\int_{t - h}^t} x^T(s) \mathcal{Q}x(s) d s \; .
\end{array}
\end{equation}
Finally, bounding the last term in \rref{equ:discus21} by $-hx^T\mathcal{Q}x$, we arrive at
\begin{equation}
\begin{array}{lll}
\mathcal{A}V_2 &\leq&  x^T Qx  - x^T(t - \tau(\rho)) Q x(t - \tau(\rho)) \\ 
&& + \displaystyle{\int_{\mathscr{B}}}\lambda(\theta,\rho) \left[ 
\int_{t - \tau(\theta)}^{t-\tau(\rho)} x^T(s)Qx(s) ds \right] d \theta \; .
\end{array}
\end{equation}
Note that, this inequality is same as the second inequality in \rref{equ:81}. Likewise,  
a similar analogy can be drawn between the last inequalities of \rref{equ:81} and \rref{equ:8}.
Therefore, a connection between the two stability results can be proved in this way. Although, Theorem \ref{Thm:1} seems to be a corollary of Theorem \ref{Thm:2}, but it is conservative due to the additional constraints (i.e., \rref{equ:3b} and \rref{equ:3c}) posed by parameter dependence. The increase in conservatism by employing fully parameter-dependent Lyapunov-Krasovskii functional is quite counter-intuitive.  We illustrate the difference in the conservatism through examples in Section 5. }

\section{Stabilization by State-feedback}

In this section, we employ the analysis presented in previous section to design a gain-scheduled feedback controller of the form
\begin{equation}
\label{cont}
\begin{array}{lll}
u(t) = K(\rho)x(t) + K_d(\rho)x(t - \tau(\rho))\;,
\end{array}
\end{equation}
where $K:\mathscr{B} \mapsto \mathbb{R}^{n_u \times n}$ and $K_d:\mathscr{B} \mapsto \mathbb{R}^{n_u \times n}$ are matrix-valued functions.   
We assume that the delay in the system dynamics is assumed to be an exactly known or measurable function of the scheduling
parameter $\rho$. While formulating the synthesis conditions for a gain-scheduled feedback, we face  a drawback of the LMI characterization posed by Theorem \ref{Thm:1} and Theorem \ref{Thm:2}. The problem arises  due to multiple product terms such as $P(\rho)A(\rho)$ and $R(\rho)A(\rho)$ in Theorem \ref{Thm:1} and $P(\rho)A(\rho)$ and $RA(\rho)$ in Theorem \ref{Thm:2}, repsectively. Therefore, we employ a slack varaible approach similar to \cite{BRIAT201510} to derive relaxed conditions. These relaxed conditions are shaped into two propositions given at the beginning of each of the following subsections.

\subsection{Stabilization Result Using Theorem \ref{Thm:1}}

\begin{proposition}
\label{prop:1}
For a given constant  $h \geq 0$, assume there exist matrix-valued functions $P:\mathscr{B} \mapsto \mathbb{S}^n_{>0}$, ${Z}:\mathscr{B} \times \mathscr{B} \mapsto \mathbb{S}^n$,  constant matrices $Q \in \mathbb{S}^n_{>0}$, $R \in \mathbb{S}^n_{>0}$ $X \in \mathbb{R}^n$, and a scalar $\gamma > 0$  such that the following 
\begin{equation}
\label{equ:20a1}
\displaystyle{\int_{\mathscr{B}}} Z(\theta,\rho) d\theta = 0
\end{equation}
holds for all $\rho \in \mathscr{B}$, and the following LMI: 
\begin{equation}
\resizebox{0.42\textwidth}{!}{$
\label{equ:middle21}
\begin{array}{lll}
\left[ 
\begin{matrix}
-\emph{Sym}[X] & P(\rho) + X^TA(\rho) &  X^TA_d(\rho) & X^TE(\rho)  \\
* & \Upsilon_{22}(\theta,\rho) & R & 0  \\ 
* & * & -Q-R & 0  \\
* & * & * & -\gamma^2 I  \\ 
* & * & * & *  \\ 
* & * & * & *  \\ 
* & * & * & *  
\end{matrix}
\right. \\ \hspace{20mm}
\left. 
\begin{matrix}
 0 & X^T & hR \\
C^T(\rho) & 0 & 0 \\
C_d^T(\rho) & 0 & 0 \\
F^T(\rho) & 0 & 0 \\
-I & 0 & 0 \\
* & -P(\rho) & -hR\\ 
 * & * & -R
\end{matrix}
\right] \prec 0 
\end{array} $}
\end{equation}
with
\begin{equation*}
\begin{array}{lll}
\Upsilon_{22}(\theta,\rho) &=&  \mu(\mathscr{B})\lambda(\theta,\rho)[P(\theta) - P(\rho)] -P(\rho)\\
&&+ Z(\theta,\rho) + \delta(\rho) Q  - R
\end{array}
\end{equation*}
holds for all $\theta,\rho \in \mathscr{B}$, where $\delta(\rho) = 1 + 2\bar{\lambda}(\rho)h$, $\bar{\lambda}(\rho) = {\int_{\mathscr{B}}}\lambda(\theta,\rho)d \theta$, and $\mu(\mathscr{B})$ is the Lebesgue measure of the set $\mathscr{B}$. Then, the system (\ref{equ:systemdef})-(\ref{equ:2}) is mean-square stable in the absence of disturbance $w$ and $u \equiv 0$. Moreover, the $\mathcal{L}_2$-gain of the map $w \mapsto z$ is at most $\gamma$.
\end{proposition}
{\it Proof:}
We start the proof by proving the feasibility of (\ref{equ:middle21}) guarantees the feasibility of (\ref{equ:middle1}). To this aim, we let (\ref{equ:middle21}) be called as ${\Upsilon}(\theta,\rho)$ and decompose it as follows:
\begin{equation*}
{\Upsilon}(\theta,\rho) = {\Upsilon}(\theta,\rho) |_{X = 0} + \mathscr{U}^T X \mathscr{V} + \mathscr{V}^T X^T \mathscr{U}\;,
\end{equation*}
where $\mathscr{U} = \left[-I_n \; A(\rho) \; A_d(\rho) \; E(\rho)\; 0_{n \times n_z} \; I_n  \; {0}_n  \right] $, and $\mathscr{V} = \left[ I_n \; {0}_{n} \; {0}_{n} \; 0_{n \times n_w} \; 0_{n \times n_z} \; {0}_{n} \; {0}_{n}  
\right]$. Then, invoking the projection lemma \cite{gahinet1994linear}, the feasibility of
${\Upsilon}(\theta,\rho) \prec 0$ implies the feasibility of the LMIs
\begin{subequations}
\begin{equation}
\label{equ:21a1}
\mathscr{N}^T_{\mathscr{U}} {\Upsilon}(\theta,\rho) |_{X = 0} \mathscr{N}_{\mathscr{U}} \prec 0 
\end{equation}
\begin{equation} 
\label{equ:21b1}
\mathscr{N}^T_{\mathscr{V}} {\Upsilon}(\theta,\rho) |_{X = 0} \mathscr{N}_{\mathscr{V}} \prec 0\;,
\end{equation}
\end{subequations}
where $\mathscr{N}_{\mathscr{U}}$ and $\mathscr{N}_{\mathscr{V}}$ are basis of the null space of $\mathscr{U}$ and $\mathscr{V}$.  
The first inequality \rref{equ:21a1} yields (\ref{equ:middle21}) and \rref{equ:21b1} yields $-R \prec 0$ for all $\rho \in \mathscr{B}$. Note that, this inequality is a relaxed form of the right bottom $1 \times 1$ block of the inequality (\ref{equ:middle21}) and is always satisfied. This concludes the proof.  \hfill $\square$

\begin{theorem}
\label{Thm:3}
For a given constant  $h \geq 0$, assume there exist matrix-valued functions $\tilde{P}:\mathscr{B} \mapsto \mathbb{S}^n_{>0}$, $\tilde{Z}:\mathscr{B} \times \mathscr{B} \mapsto \mathbb{S}^n$, $Y:\mathscr{B} \mapsto \mathbb{R}^{n_u \times n}$, $Y_d:\mathscr{B} \mapsto \mathbb{R}^{n_u \times n}$,  constant matrices $\tilde{Q} \in \mathbb{S}^n_{>0}$, $R \in \mathbb{S}^n_{>0}$, $\tilde{X} \in \mathbb{R}^n$, and a  scalar $\gamma > 0$  such that the following equality
\begin{equation}
\label{equ:27a1}
\displaystyle{\int_{\mathscr{B}}} \tilde{Z}(\theta,\rho) d\theta = 0
\end{equation}
holds for all $\rho \in \mathscr{B}$, and the following LMI:
\begin{equation}
\resizebox{0.5\textwidth}{!}{$
\label{equ:middle31}
\begin{array}{lll}
\left[ 
\begin{matrix}
-\emph{Sym}[\tilde{X}] & \tilde{\Upsilon}_{12}&  \tilde{\Upsilon}_{13} & E(\rho) &0 & \tilde{X} & h\tilde{R}  \\
* & \tilde{\Upsilon}_{22} & \tilde{R} & 0 & \tilde{\Upsilon}_{25} & 0 & 0  \\ 
* & * & \tilde{\Upsilon}_{33} & 0 & \tilde{\Upsilon}_{35} & 0 & 0 \\
* & * & * & -\gamma^2 I & F^T(\rho) & 0 & 0 \\ 
* & * & * & * & -I & 0 & 0 \\ 
* & * & * & * & * & -\tilde{P}(\rho) & -h\tilde{R} \\ 
* & * & * & *  &  * & * & -\tilde{R}
\end{matrix}
\right] \prec 0 
\end{array} $}
\end{equation}
with
\begin{equation*}
\fontsize{9}{9} \selectfont
\begin{array}{lll}
\tilde{\Upsilon}_{12} &=& \tilde{P}(\rho) + A(\rho)\tilde{X} + B(\rho)Y(\rho) \\ 
\tilde{\Upsilon}_{13} &=& A_d(\rho)\tilde{X} + B(\rho)Y_d(\rho) \\
\Upsilon_{22} &=&  \mu(\mathscr{B})\lambda(\theta,\rho)[\tilde{P}(\theta) - \tilde{P}(\rho)] -\tilde{P}(\rho)\\
&&+ \tilde{Z}(\theta,\rho) + \delta(\rho) \tilde{Q} - \tilde{R} \\ 
\tilde{\Upsilon}_{25} &=& [C(\rho)\tilde{X} + D(\rho)Y(\rho)]^T, \
\tilde{\Upsilon}_{33} = -\tilde{Q}-\tilde{R} \\  
\tilde{\Upsilon}_{35} &=& [C_d(\rho)\tilde{X} + D(\rho)Y_d(\rho)]^T
\end{array}
\end{equation*}
holds for all $\theta,\rho \in \mathscr{B}$, where $\delta(\rho) = 1 + 2\bar{\lambda}(\rho)h$, $\bar{\lambda}(\rho) = {\int_{\mathscr{B}}}\lambda(\theta,\rho)d \theta$, and $\mu(\mathscr{B})$ is the Lebesgue measure of the set $\mathscr{B}$. Then, the closed-loop system (\ref{equ:systemdef})-(\ref{equ:2})-(\ref{cont}) is mean-square stable in the absence of disturbance $w$ and the $\mathcal{L}_2$-gain of the map $w \mapsto z$ is at most $\gamma$. Moreover, the gains of the controller are given as: 
\begin{equation*}
\begin{array}{lll}
K(\rho) = Y(\rho)\tilde{X}^{-1}, \; K_d(\rho) = Y_d(\rho)\tilde{X}^{-1}.
\end{array}
\end{equation*}
\end{theorem}
{\it Proof:}
First, substitute the closed-loop system matrices
\begin{equation*}
\begin{array}{lll}
A(\rho) &\leftarrow& A_{cl}(\rho) := A(\rho) + B(\rho)K(\rho) \\ 
A_d(\rho) &\leftarrow& A_{dcl}(\rho) := A_d(\rho) + B(\rho)K_d(\rho) \\ 
C(\rho) &\leftarrow& C_{cl}(\rho) := C(\rho) + D(\rho)K(\rho) \\ 
C_d(\rho) &\leftarrow& C_{dcl}(\rho) := C_d(\rho) + D(\rho)K_d(\rho) \\ 
\end{array}
\end{equation*}
into the LMI \rref{equ:middle21} and then perform a congruence transformation with respect to $diag(\tilde{X},\tilde{X},\tilde{X},I_{n_w},I_{n_z},\tilde{X},\tilde{X})$, where $\tilde{X} = X^{-1}$. Employing the following linearizing change of variables 
$\tilde{P}(\rho) = \tilde{X}^T P(\rho) \tilde{X}$,$\tilde{P}(\theta) = \tilde{X}^T P(\theta) \tilde{X}$, $\tilde{Q} = \tilde{X}^T Q \tilde{X}$, $\tilde{R}(\rho) = \tilde{X}^T R \tilde{X}$, $\tilde{Z}(\theta,\rho) = \tilde{X}^T Z(\theta,\rho) \tilde{X}$, $Y(\rho) = K(\rho)\tilde{X}$, and $Y_d(\rho) = K_d(\rho)\tilde{X}$ 
yield the result. \hfill $\square$ 

\subsection{Stabilization Result Using Theorem \ref{Thm:2}}

\begin{proposition}
\label{prop:2}
For a given constant  $h \geq 0$, assume there exist matrix-valued functions $P:\mathscr{B} \mapsto \mathbb{S}^n_{>0}$, $Q:\mathscr{B} \mapsto \mathbb{S}^n_{>0}$, $R:\mathscr{B} \mapsto \mathbb{S}^n_{>0}$, $\mathcal{Q}:\mathscr{B} \mapsto \mathbb{S}^n_{>0}$, ${Z}:\mathscr{B} \times \mathscr{B} \mapsto \mathbb{S}^n$, a constant matrix $X \in \mathbb{R}^n$, and the  scalars $\hat{\lambda} > 0$, and $\gamma > 0$  such that following 
\begin{subequations}
\small
\begin{equation}
\label{equ:20a}
\displaystyle{\int_{\mathscr{B}}} Z(\theta,\rho) d\theta = 0
\end{equation}
\begin{equation}
\label{equ:20b}
\displaystyle{\int_{\mathscr{B}}} \lambda(\theta,\rho)Q(\theta) d \theta \leq \mathcal{Q}(\rho) 
\end{equation}
\begin{equation}
\label{equ:20c}
\displaystyle{\int_{\mathscr{B}}} \lambda(\theta,\rho) R(\theta) d \theta \leq \hat{\lambda} R(\rho)
\end{equation} 
\end{subequations}
hold for all $\rho \in \mathscr{B}$, and the following LMI: 
\begin{equation}
\resizebox{0.42\textwidth}{!}{$
\label{equ:middle2}
\begin{array}{lll}
\left[ 
\begin{matrix}
-\emph{Sym}[X] & P(\rho) + X^TA(\rho) &  X^TA_d(\rho) & X^TE(\rho)  \\
* & \Upsilon_{22}(\theta,\rho) & R(\rho) & 0  \\ 
* & * & -Q(\rho)-R(\rho) & 0  \\
* & * & * & -\gamma^2 I  \\ 
* & * & * & *  \\ 
* & * & * & *  \\ 
* & * & * & *  
\end{matrix}
\right. \\ \hspace{20mm}
\left. 
\begin{matrix}
 0 & X^T & \sqrt{\epsilon}R(\rho) \\
C^T(\rho) & 0 & 0 \\
C_d^T(\rho) & 0 & 0 \\
F^T(\rho) & 0 & 0 \\
-I & 0 & 0 \\
* & -P(\rho) & -\sqrt{\epsilon}R(\rho) \\ 
 * & * & -R(\rho)
\end{matrix}
\right] \prec 0 
\end{array} $}
\end{equation}
with
\begin{equation*}
\begin{array}{lll}
\Upsilon_{22}(\theta,\rho) &=&  \mu(\mathscr{B})\lambda(\theta,\rho)[P(\theta) - P(\rho)] -P(\rho)\\
&&+ Z(\theta,\rho) + Q(\rho) + h \mathcal{Q} - R(\rho)
\end{array}
\end{equation*}
holds for all $\theta,\rho \in \mathscr{B}$, where $\epsilon = {h^2} + \hat{\lambda}\frac{h^3}{2}$ and $\mu(\mathscr{B})$ is the Lebesgue measure of the set $\mathscr{B}$. Then, the system (\ref{equ:systemdef})-(\ref{equ:2}) is mean-square stable in the absence of disturbance $w$ and $u \equiv 0$. Moreover, the $\mathcal{L}_2$-gain of the map $w \mapsto z$ is at most $\gamma$.
\end{proposition}
{\it Proof:}
The proof follows the same arguments as the proof of Proposition \ref{prop:1}. \hfill $\square$

\begin{theorem}
\label{Thm:4}
For a given constant  $h \geq 0$, assume there exist matrix-valued functions $\tilde{P}:\mathscr{B} \mapsto \mathbb{S}^n_{>0}$, $\tilde{Q}:\mathscr{B} \mapsto \mathbb{S}^n_{>0}$, $R:\mathscr{B} \mapsto \mathbb{S}^n_{>0}$, $\tilde{\mathcal{Q}}:\mathscr{B} \mapsto \mathbb{S}^n_{>0}$, $\tilde{Z}:\mathscr{B} \times \mathscr{B} \mapsto \mathbb{S}^n$, $Y:\mathscr{B} \mapsto \mathbb{R}^{n_u \times n}$, $Y_d:\mathscr{B} \mapsto \mathbb{R}^{n_u \times n}$, a constant matrix $\tilde{X} \in \mathbb{R}^n$, and  scalars $\hat{\lambda} > 0$, and $\gamma > 0$  such that following 
\begin{subequations}
\fontsize{9}{9} \selectfont
\begin{equation}
\label{equ:27a}
\displaystyle{\int_{\mathscr{B}}} \tilde{Z}(\theta,\rho) d\theta = 0
\end{equation}
\begin{equation}
\label{equ:27b}
\displaystyle{\int_{\mathscr{B}}} \lambda(\theta,\rho)\tilde{Q}(\theta) d \theta \leq \tilde{\mathcal{Q}}(\rho) 
\end{equation}
\begin{equation}
\label{equ:27c}
\displaystyle{\int_{\mathscr{B}}} \lambda(\theta,\rho) \tilde{R}(\theta) d \theta \leq \hat{\lambda} \tilde{R}(\rho)
\end{equation} 
\end{subequations}
hold for all $\rho \in \mathscr{B}$, and the following LMI:
\begin{equation}
\resizebox{0.5\textwidth}{!}{$
\label{equ:middle3}
\begin{array}{lll}
\left[ 
\begin{matrix}
-\emph{Sym}[\tilde{X}] & \tilde{\Upsilon}_{12}&  \tilde{\Upsilon}_{13} & E(\rho) &0 & \tilde{X} & \sqrt{\epsilon}\tilde{R}(\rho)  \\
* & \tilde{\Upsilon}_{22} & \tilde{R}(\rho) & 0 & \tilde{\Upsilon}_{25} & 0 & 0  \\ 
* & * & \tilde{\Upsilon}_{33} & 0 & \tilde{\Upsilon}_{35} & 0 & 0 \\
* & * & * & -\gamma^2 I & F^T(\rho) & 0 & 0 \\ 
* & * & * & * & -I & 0 & 0 \\ 
* & * & * & * & * & -\tilde{P}(\rho) & -\sqrt{\epsilon}\tilde{R}(\rho) \\ 
* & * & * & *  &  * & * & -\tilde{R}(\rho)
\end{matrix}
\right] \prec 0 
\end{array} $}
\end{equation}
with
\begin{equation*}
\fontsize{9}{9} \selectfont
\begin{array}{lll}
\tilde{\Upsilon}_{12} &=& \tilde{P}(\rho) + A(\rho)\tilde{X} + B(\rho)Y(\rho) \\ 
\tilde{\Upsilon}_{13} &=& A_d(\rho)\tilde{X} + B(\rho)Y_d(\rho) \\
\Upsilon_{22} &=&  \mu(\mathscr{B})\lambda(\theta,\rho)[\tilde{P}(\theta) - \tilde{P}(\rho)] -\tilde{P}(\rho)\\
&&+ \tilde{Z}(\theta,\rho) + \tilde{Q}(\rho) + h \tilde{\mathcal{Q}} - \tilde{R}(\rho) \\ 
\tilde{\Upsilon}_{25} &=& [C(\rho)\tilde{X} + D(\rho)Y(\rho)]^T, \
\tilde{\Upsilon}_{33} = -\tilde{Q}(\rho)-\tilde{R}(\rho) \\  
\tilde{\Upsilon}_{35} &=& [C_d(\rho)\tilde{X} + D(\rho)Y_d(\rho)]^T
\end{array}
\end{equation*}
holds for all $\theta,\rho \in \mathscr{B}$, where $\epsilon = {h^2} + \hat{\lambda}\frac{h^3}{2}$, and $\mu(\mathscr{B})$ is the Lebesgue measure of the set $\mathscr{B}$. Then, the closed-loop system (\ref{equ:systemdef})-(\ref{equ:2})-(\ref{cont}) is mean-square stable in the absence of disturbance $w$ and the $\mathcal{L}_2$-gain of the map $w \mapsto z$ is at most $\gamma$. Moreover, the gains of the controller are given as: 
\begin{equation*}
\begin{array}{lll}
K(\rho) = Y(\rho)\tilde{X}^{-1}, \; K_d(\rho) = Y_d(\rho)\tilde{X}^{-1}.
\end{array}
\end{equation*}
\end{theorem}
{\it Proof:}
The proof follows the same arguments as for the proof of Theorem \ref{Thm:3} by employing following linearizing change of variables 
$\tilde{P}(\rho) = \tilde{X}^T P(\rho) \tilde{X}$,$\tilde{P}(\theta) = \tilde{X}^T P(\theta) \tilde{X}$, $\tilde{Q}(\rho) = \tilde{X}^T Q(\rho) \tilde{X}$, $\tilde{R}(\rho) = \tilde{X}^T R(\rho) \tilde{X}$, $\tilde{Z}(\theta,\rho) = \tilde{X}^T Z(\theta,\rho) \tilde{X}$, $\tilde{\mathcal{Q}}(\rho) = \tilde{X}^T \mathcal{Q}(\rho) \tilde{X}$, $Y(\rho) = K(\rho)\tilde{X}$, and $Y_d(\rho) = K_d(\rho)\tilde{X}$. \hfill $\square$ 
\begin{remark}
The LMIs formulated in the above results yield intractable infinite-dimensional semi-definite programs. To make them tractable, several methods are available, such as sum of squares programming \cite{scherer2006matrix},  gridding approach \cite[Appendix C]{BriatBook}, etc. See \cite{briat2018stability} for their implementation and computational aspects. These methods will result in an approximate finite-dimensional
semi-definite program which can then be solved using standard solvers such as SeDuMi \cite{sturm1999using}. 
\end{remark}
\begin{remark}
The integral equality constraints on $Z$ and $\tilde{Z}$ in the main results can be easily implemented using YALMIP \cite{lofberg2004yalmip} as they are simply equality constraints on the coefficients on the matrix polynomials $Z$ and $\tilde{Z}$.  
\end{remark}

\section{Illustrations}
\subsection{Stability Analysis}
\begin{figure}
\centering
\includegraphics[width = \linewidth]{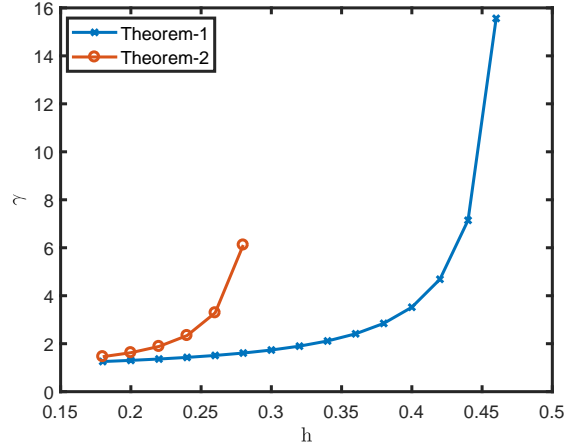}
\caption{Minimum $\gamma$ achieved against the upper bound $h$ for fixed $\lambda_0 = 10$ using Theorem \ref{Thm:1} and \ref{Thm:2}.}
\label{fig:1}
\end{figure}

\begin{figure}
\centering
\includegraphics[width = \linewidth]{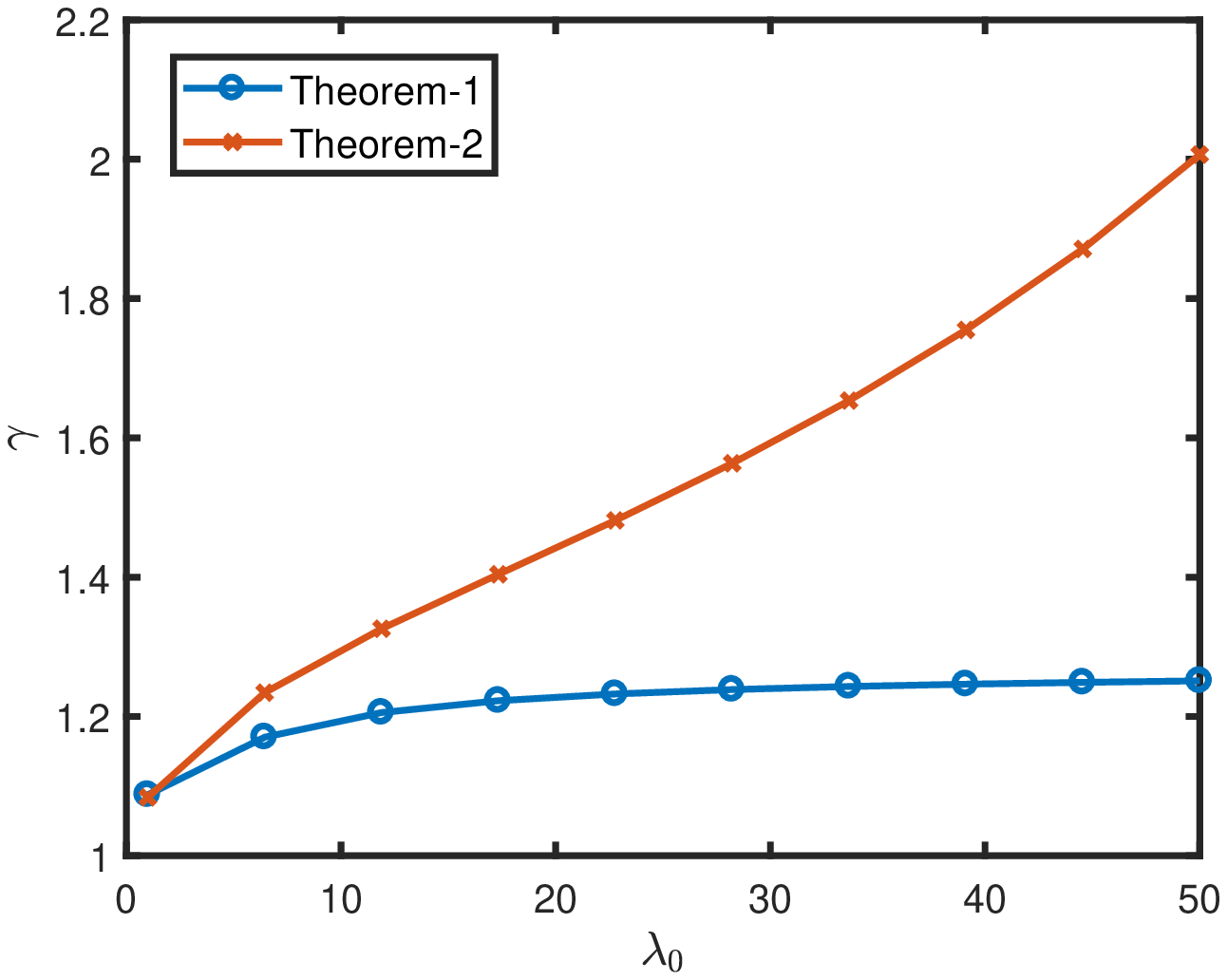}
\caption{Minimum $\gamma$ achieved against the intensities $\lambda$ for fixed $h = 0.15$ using Theorem \ref{Thm:1} and \ref{Thm:2}.}
\label{fig:2}
\end{figure}

Consider the system \rref{equ:systemdef}-\rref{equ:2} without input (i.e., $u \equiv 0$) and with following matrices  
\begin{equation*}
\begin{array}{lll}
A(\rho) = \left[ \begin{matrix}
0 & 1 \\ -2-\rho & 1
\end{matrix}
\right], \; 
A_d(\rho) = \left[
\begin{matrix}
-1 & 0 \\ -1-\rho & -1
\end{matrix}
\right], \\[4mm] 
C = \left[
\begin{matrix}
1 & 0
\end{matrix}
\right], \; 
C_d = C,  \; 
E  = C^T, \; 
F = 0. 
\end{array}
\end{equation*}
The stochastic parameter $\rho$ takes the value in $\mathscr{B} =[0,\ 1]$. For the comparison purposes, we solved the LMIs in Theorem \ref{Thm:1} and Theorem \ref{Thm:2} via a gridding approach using fifty points. First-order polynomials have been employed to approximate the unknown matrix-valued functions appearing in \rref{equ:middle1} and \rref{equ:middle}. We study two cases: Case I) fix the intensity $\lambda(\theta,\rho)=  10$ and minimize the performance index $\gamma$ against the upper bounds $h$ on time-delays, Case II) fix the upper bound $h=0.15$ and
 minimize $\gamma$ against the intensities $\lambda_0$. The results for Case I and Case II are illustrated in Fig. \ref{fig:1} and Fig. \ref{fig:2}, respectively. The figures demonstrate that given the same scenario, Theorem \ref{Thm:1} yields smaller values of $\gamma$ as compared to Theorem \ref{Thm:2}. Hence, the result of Theorem \ref{Thm:1} is less conservative as compared to Theorem \ref{Thm:2}.
\subsection{Stabilization by State-Feedback}
\begin{figure}
\centering
\includegraphics[width = \linewidth]{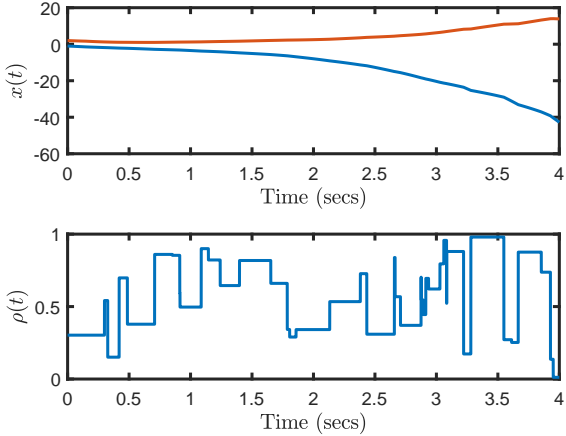}
\caption{Evolution of the states of the open-loop system (top) and a typical trajectory for the parameter (bottom).}
\label{fig:3}
\end{figure}

\begin{figure}
\centering
\includegraphics[width = \linewidth]{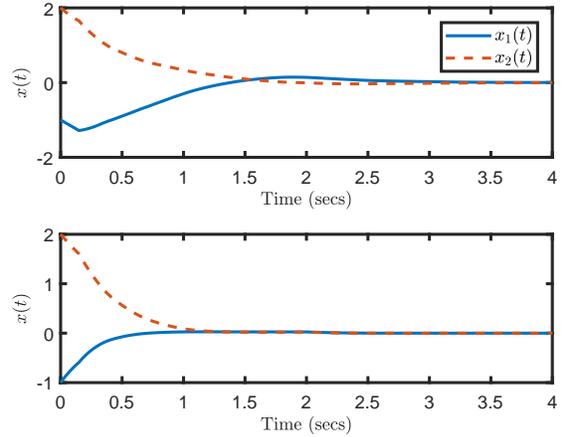}
\caption{Evolution of the states of the closed-loop systems with controllers gains from Theorem \ref{Thm:3} (top) and Theorem \ref{Thm:4} (bottom)  subject to disturbance $w(t) = H(t)-H(t-2)$ and stochastic time-delay $\tau(\rho) = 0.5\sin(\rho)$.}
\label{fig:4}
\end{figure}

\begin{figure}
\centering
\includegraphics[width = \linewidth]{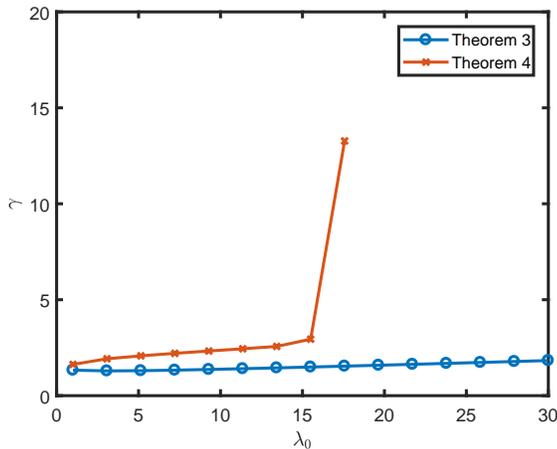}
\caption{Minimum $\gamma$ achieved against the intensities $\lambda_0$ for fixed $h=0.5$ using Theorem \ref{Thm:3} and  \ref{Thm:4}.}
\label{fig:5}
\end{figure}

This example compares Theorem \ref{Thm:3} and Theorem \ref{Thm:4} in terms of the conservatism. Consider the system \rref{equ:systemdef}-\rref{equ:2} with the  following matrices: 
\begin{equation*}
\resizebox{0.48\textwidth}{!}{$
\begin{array}{lll}
A(\rho) = \left[
\begin{matrix}
2-\rho& -0.5-0.5\rho \\ -1 & -2+0.1\rho
\end{matrix}
\right], \; 
  A_d(\rho) = \left[
  \begin{matrix}
  -1 & 0 \\ 0.05-0.45\rho & -1
  \end{matrix}
  \right], \\
  B = \left[ 
\begin{matrix}
1  \\ 0 
\end{matrix}  
  \right], \; 
  E = \left[ 
\begin{matrix}
0.1  \\ 0.1
\end{matrix}  
  \right], \; 
  C = \left[
\begin{matrix}
0 & 1
\end{matrix}  
  \right], \; 
  C_d = C, \;  
  D = 1,\; F = 0.    
\end{array} $}
\end{equation*}
We choose $h = 0.5$ and $\mathscr{B} = [0,1]$. The open-loop response in Fig. \ref{fig:3} (above) illustrates unstable behavior subject to a typical realization of stochastic parameter provided at the bottom of the same figure. 
We solved the parameter-dependent LMIs in Theorem \ref{Thm:3} and Theorem \ref{Thm:4}  by first assuming parameter dependence as polynomial functions of $\rho$, and then employing the gridding approach. We employ first-order polynomials along with fifty gridding points to compute the unknown matrices in the Theorem \ref{Thm:3} and Theorem \ref{Thm:4}. Once the semi-definite programs are feasible, we compute the state-feedback controllers. The computed controllers from Theorem \ref{Thm:3} and Theorem \ref{Thm:4} are denoted by superscript `3' and `4', respectively, and are given below   
\begin{equation*}
\fontsize{9}{9} \selectfont
\begin{array}{lll}
K^3(\rho) &=&
\displaystyle{\frac{1}{\Delta}}\left[ 
\begin{matrix}
11.446\rho - 15.689 &  6.18\rho - 1.1362 \end{matrix} \right] \\[4mm] 
K_d^3(\rho) &=&
\displaystyle{\frac{1}{\Delta}} \left[ 
\begin{matrix}
- 0.99326\rho - 12.622& 0.15588\rho + 3.0325
\end{matrix}
\right]\;,
\end{array}
\end{equation*} 
where $\Delta = 13.678$, and 
\begin{equation*}
\fontsize{9}{9} \selectfont
\begin{array}{lll}
K^4(\rho) &=&
\displaystyle{\frac{1}{\bar{\Delta}}}\left[ 
\begin{matrix}
75.681\rho - 386.55& 33.006\rho + 49.741 \end{matrix} \right] \\[4mm] 
K_d^4(\rho) &=&
\displaystyle{\frac{1}{\bar{\Delta}}} \left[ 
\begin{matrix}
1.6124\rho + 6.1874& 0.36509\rho - 1.2883
\end{matrix}
\right]\;,
\end{array}
\end{equation*} 
where $\bar{\Delta} = 76.136$. 
The stabilizing closed-loop responses for an initial condition $[-1,2]$ subject to disturbance $w(t) = H(t)-H(t-2)$,
where $H(t)$ is the Heavyside step function and stochastic time-delay $\tau(\rho) = 0.5\sin(\rho)$ for $\rho \in \mathscr{B}$ are demonstrated in Fig. \ref{fig:4}. The conservatism in Theorem \ref{Thm:4} reflects by the fact that, the LMIs \rref{equ:middle3} are feasible only for $\lambda_0 \leq 17$ with $\hat{\lambda} = \lambda_0+\delta_\lambda$, where $\delta_\lambda = 0.005$. On the other hand, the LMIs \rref{equ:middle31} in Theorem \ref{Thm:3} yields feasible solutions for larger intensities while yielding smaller values of $\gamma$ as depicted in Fig. \ref{fig:5}.

\section{Concluding Remarks}
Stability analysis and state-feedback stabilization of LPV time-delay systems with piecewise constant parameters subject to spontaneous Poissonian jumps and stochastic delays are discussed. These systems generalize the framework of MJLS with time-delay associated with finite/infinite countable sets to bounded uncountable sets. We demonstrate that the parameterization of stochastic Lyapunov-Krasovskii functional critically affects the conservatism in the analysis. Although, considering parameter dependence in every decision variable of Lyapunov-Krasovskii functionals seems to reduce the conservatism in general theory, but in this case, the conservatism is increased owing to additional constraints posed by parameter dependence.      

Future extensions of our work include: (i) designing a dynamic output feedback controller in a convex non-conservative way and its control-theoretic applications, (ii) deriving stability analysis by considering stochastic delays in input and output of the system, and (iii) employing a recent integral-based Wirtinger's inequality  \cite{seuret2013wirtinger} to upper bound the integrals in Lyapunov-Krasovskii functionals to further reduce the conservatism. These extensions will be reported elsewhere.

\begin{ack}                             
The author would like to acknowledge fruitful discussions with Dr. Corentin Briat. 
\end{ack}

\bibliographystyle{plain}        
\bibliography{autosam}

\begin{thebibliography}{10}

\bibitem{boukas2012deterministic}
El-Kebir Boukas and Zi-Kuan Liu.
\newblock {\em Deterministic and stochastic time-delay systems}.
\newblock Springer Science \& Business Media, 2012.

\bibitem{BriatBook}
Corentin Briat.
\newblock {\em Linear Parameter-Varying and Time-Delay Systems -- Analysis,
  Observation, Filtering \& Control}.
\newblock Springer-Verlag, Berlin-Heidelberg, 2015.

\bibitem{BRIAT201510}
Corentin Briat.
\newblock Stability analysis and control of a class of {LPV} systems with
  piecewise constant parameters.
\newblock {\em Systems \& Control Letters}, 82:10 -- 17, 2015.

\bibitem{briat2018stability}
Corentin Briat.
\newblock Stability analysis and state-feedback control of {LPV} systems with
  piecewise constant parameters subject to spontaneous poissonian jumps.
\newblock {\em IEEE Control Systems Letters}, 2(2):230--235, 2018.

\bibitem{briat2007lft}
Corentin Briat, Olivier Sename, and Jean-Fran{\c{c}}ois Lafay.
\newblock A {LFT/$H_\infty$} state feedback design for linear parameter varying
  time delay systems.
\newblock In {\em Proceedings of the European Control Conference}, pages
  4882--4888, 2007.

\bibitem{briat2008parameter}
Corentin Briat, Olivier Sename, and Jean-Fran{\c{c}}ois Lafay.
\newblock Parameter dependent state-feedback control of {LPV} time delay
  systems with time varying delays using a projection approach.
\newblock {\em IFAC Proceedings Volumes}, 41(2):4946--4951, 2008.

\bibitem{briat2010memory}
Corentin Briat, Olivier Sename, and Jean-Fran{\c{c}}ois Lafay.
\newblock Memory-resilient gain-scheduled state-feedback control of uncertain
  {LTI/LPV} systems with time-varying delays.
\newblock {\em Systems \& Control Letters}, 59(8):451--459, 2010.

\bibitem{chiasson2007applications}
John Chiasson and Jean~Jacques Loiseau.
\newblock {\em Applications of Time Delay Systems}.
\newblock Springer-Verlag, Berlin-Heidelberg, 2007.

\bibitem{davis2018markov}
Mark~HA Davis.
\newblock {\em Markov models \& optimization}.
\newblock Routledge, 2018.

\bibitem{de2000output}
Daniela~Pucci De~Farias, Jos{\'e}~Claudio Geromel, Jo{\~a}o~BR Do~Val, and
  Oswaldo Luiz~V Costa.
\newblock Output feedback control of {M}arkov jump linear systems in
  continuous-time.
\newblock {\em IEEE Transactions on Automatic Control}, 45(5):944--949, 2000.

\bibitem{gahinet1994linear}
Pascal Gahinet and Pierre Apkarian.
\newblock A linear matrix inequality approach to ${H_\infty}$ control.
\newblock {\em International Journal of Robust and Nonlinear Control},
  4(4):421--448, 1994.

\bibitem{gilbert2010polynomial}
Wilfried Gilbert, Didier Henrion, Jacques Bernussou, and David Boyer.
\newblock Polynomial {LPV} synthesis applied to turbofan engines.
\newblock {\em Control Engineering Practice}, 18(9):1077--1083, 2010.

\bibitem{gu2003stability}
Keqin Gu, Vladimir~L Kharitonov, and Jie Chen.
\newblock {\em Stability of Time-Delay Systems}.
\newblock Birkh\"{a}user, Basel, 2003.

\bibitem{hosoe2018robust}
Yohei Hosoe, Tomomichi Hagiwara, and Dimitri Peaucelle.
\newblock Robust stability analysis and state feedback synthesis for
  discrete-time systems characterized by random polytopes.
\newblock {\em IEEE Transactions on Automatic Control}, 63(2):556--562, 2018.

\bibitem{kajiwara1999lpv}
Hiroyuki Kajiwara, Pierre Apkarian, and Pascal Gahinet.
\newblock {LPV} techniques for control of an inverted pendulum.
\newblock {\em IEEE Control Systems Magazine}, 19(1):44--54, 1999.

\bibitem{lofberg2004yalmip}
Johan L{\"o}fberg.
\newblock {YALMIP}: A toolbox for modeling and optimization in {MATLAB}.
\newblock In {\em Proceedings of the CACSD Conference}, volume~3. Taipei,
  Taiwan, 2004.

\bibitem{mohammadpour2012control}
Javad Mohammadpour and Carsten~W Scherer.
\newblock {\em Control of Linear Parameter Varying Systems with Applications}.
\newblock Springer-Verlag, New York, 2012.

\bibitem{niculescu2001delay}
Silviu-Iulian Niculescu.
\newblock {\em Delay Effects on Stability: A Robust Control Approach}.
\newblock Springer-Verlag, London, 2001.

\bibitem{peet2009positive}
Matthew~M Peet, Antonis Papachristodoulou, and Sanjay Lall.
\newblock Positive forms and stability of linear time-delay systems.
\newblock {\em SIAM Journal on Control and Optimization}, 47(6):3237--3258,
  2009.

\bibitem{scherer2006matrix}
Carsten~W Scherer and Camile~WJ Hol.
\newblock Matrix sum-of-squares relaxations for robust semi-definite programs.
\newblock {\em Mathematical programming}, 107(1-2):189--211, 2006.

\bibitem{sename2013robust}
Olivier Sename, Peter Gaspar, and J{\'o}zsef Bokor.
\newblock {\em Robust control and linear parameter varying approaches:
  application to vehicle dynamics}, volume 437.
\newblock Springer, 2013.

\bibitem{seuret2013wirtinger}
Alexandre Seuret and Fr{\'e}d{\'e}ric Gouaisbaut.
\newblock Wirtinger-based integral inequality: Application to time-delay
  systems.
\newblock {\em Automatica}, 49(9):2860--2866, 2013.

\bibitem{shin2000h}
Jong-Yeob Shin, Gary~J Balas, and Andrew~K Packard.
\newblock {$H_\infty$} control of the {V132 X-38} lateral-directional axis.
\newblock In {\em Proceedings of the 2000 American Control Conference. ACC
  (IEEE Cat. No. 00CH36334)}, volume~3, pages 1862--1866. IEEE, 2000.

\bibitem{sturm1999using}
Jos~F Sturm.
\newblock Using {SeDuMi} 1.02, a {MATLAB} toolbox for optimization over
  symmetric cones.
\newblock {\em Optimization methods and software}, 11(1-4):625--653, 1999.

\bibitem{teel2014stability}
Andrew~R Teel, Anantharaman Subbaraman, and Antonino Sferlazza.
\newblock Stability analysis for stochastic hybrid systems: {A} survey.
\newblock {\em Automatica}, 50(10):2435--2456, 2014.

\bibitem{todorov2008output}
Marcos~G Todorov and Marcelo~D Fragoso.
\newblock Output feedback {$H_\infty$} control of continuous-time infinite
  {M}arkovian jump linear systems via {LMI} methods.
\newblock {\em SIAM Journal on Control and Optimization}, 47(2):950--974, 2008.

\bibitem{wu2001lpv}
Fen Wu and Karolos~M Grigoriadis.
\newblock {LPV} systems with parameter-varying time delays: analysis and
  control.
\newblock {\em Automatica}, 37(2):221--229, 2001.

\bibitem{zakwan2019poisson}
M.~{Zakwan}.
\newblock Dynamic {$L_{\text{2}}$} output feedback stabilization of {LPV}
  systems with piecewise constant parameters subject to spontaneous poissonian
  jumps.
\newblock {\em IEEE Control Systems Letters}, 4(2):408--413, 2020.

\bibitem{zakwan2020distributedj}
Muhammad Zakwan and Saeed Ahmed.
\newblock Distributed output feedback control of decomposable {LPV} systems
  with delay and switching topology: Application to consensus problem in
  multi-agent systems.
\newblock {\em International Journal of Control.
  \emph{doi:10.1080/00207179.2019.1710257}}, 2020.

\bibitem{zakwan2020Dwell}
Muhammad Zakwan and Saeed Ahmed.
\newblock Dwell-time based stability analysis and $\mathcal{L}_2$ control of
  {LPV} systems with piecewise constant parameters and delay.
\newblock {\em arXiv preprint arXiv:2002.12102}, 2020.

\bibitem{zhang2005delay}
Feng Zhang and Karolos~M Grigoriadis.
\newblock Delay-dependent stability analysis and ${H_\infty}$control for
  state-delayed {LPV} system.
\newblock In {\em Proceedings of the 2005 IEEE International Symposium on,
  Mediterrean Conference on Control and Automation Intelligent Control, 2005.},
  pages 1532--1537. IEEE, 2005.

\bibitem{zhang2002stability}
Xiping Zhang, Panagiotis Tsiotras, and Carl Knospe.
\newblock Stability analysis of {LPV} time-delayed systems.
\newblock {\em International Journal of Control}, 75(7):538--558, 2002.

\end{thebibliography}

\end{document}